\documentclass[aps,twocolumn,floatfix,superscriptaddress]{revtex4}
\usepackage{amssymb}
\usepackage{graphicx}

\begin{document}

\title{
Topological Properties of Tensor Network States\\
From Their Local Gauge and Local Symmetry Structures
}

\author{Brian Swingle}
\affiliation{Department of Physics, Massachusetts Institute of
Technology, Cambridge, Massachusetts 02139, USA }

\author{Xiao-Gang Wen}
\affiliation{Department of Physics, Massachusetts Institute of
Technology, Cambridge, Massachusetts 02139, USA }

\begin{abstract}
Tensor network states are capable of describing many-body systems with complex
quantum entanglement, including systems with
non-trivial topological order. In this paper, we study 
methods to calculate the topological properties of a tensor network 
state from the tensors that form the state.
Motivated by the concepts of gauge group and projective
symmetry group in the slave-particle/projective
construction, and by the low-dimensional gauge-like
symmetries of some exactly solvable Hamiltonians, we study
the $d$-dimensional gauge structure and the $d$-dimensional
symmetry structure of a tensor network state, where $d\leq
d_\text{space}$ with $d_\text{space}$ the dimension of
space.  The $d$-dimensional gauge structure and
$d$-dimensional symmetry structure allow us to calculate the
string operators and $d$-brane operators of the tensor
network state. This in turn allows us to calculate many
topological properties of the tensor network state, such as
ground state degeneracy and quasiparticle statistics.
\end{abstract}

\maketitle

\section{Introduction}

Topological order\cite{Wtop,Wrig,Wtoprev} is a new kind of
ordering in many-body quantum states. It represents new
states of quantum matter beyond the symmetry breaking states
characterized by Landau symmetry breaking theory.  This
new kind of order and the new states of matter associated with it open up a whole new
research direction in condensed matter
physics.\cite{KL8795,WWZcsp,MR9162,Wnab,RS9173,Wsrvb,MLB9964,MSP0202,K032,MS0312,FNS0428,LWsta,LWstrnet,C0502,NSS0883}
Intuitively, topological order corresponds to a pattern of long
range quantum entanglement in the ground state.

Traditional mean-field approaches to calculating the $T=0$ phase
diagram of a quantum system are based on direct product
states which have no long range quantum entanglement. These approaches seem to exclude topological states from the very
beginning.  To study a phase diagram that may contain
topological states, we have to find a way write down states
with non-trivial long range entanglement. The tensor network
state (TNS)\cite{GMN0391,NMG0415,M0460,VC0466}
method is one way to produce states with long range
entanglement.\cite{AV0804,GLS0918,BAV0919}  We can
develop a mean-field approach for topologically ordered states
based on TNS.\cite{LN0701,JOV0802,GLWtergV,JWX0803,GWtefr,OV0903}

Physically, topological order is characterized and defined
through measurable quantities, such as ground state
degeneracy on a torus or other compact
space\cite{Wtop,WNtop} and fractionalized quantum numbers
and statistics of quasiparticles \cite{L8395,ASW8422}.
If we claim that TNS can describe topological
states, then it is natural to ask how to calculate those
measurable topological quantum numbers from the TNS representation.  In this
paper, we will address this question.  We will show how to
calculate measurable topological properties of a TNS.

Our approach is motivated by the gauge group and projective
symmetry group in the slave-particle/projective construction
\cite{BA8880,Wqoslpub} and by the low-dimensional gauge-like
symmetries of some exactly soluble
Hamiltonians.\cite{NO0616} We will introduce the
local gauge and local symmetry structures of a TNS from the
tensors in the TNS. This allows us to obtain the string
operators (or more generally, the $d$-brane operators) of
the TNS, which in turn allows us to calculate the
topological properties of the TNS.

\section{TNS, ideal wave functions, and ideal Hamiltonians}

\begin{figure}
\begin{center}
\includegraphics[width=2.7in]{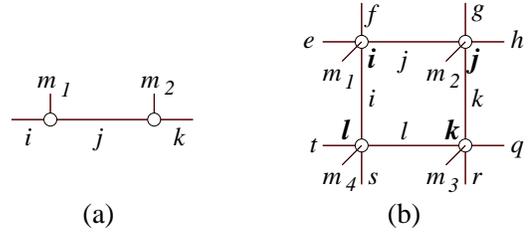}
\end{center}
%Fig 1
\caption{Tensor-network -- a graphical representation of the
tensor network wave function \eq{TNS}, (a) on a 1D chain  or (b) on
a 2D square lattice.  The indices on the links are summed over. }
\label{tps}
\end{figure}

What is a TNS? As an example, let us consider a TNS on
a square lattice, where the physical states living on
each vertex $\v i$ are labeled by $m_{\v i}$.
The TNS is defined by the following expression for the many-body wave
function $\Psi(\{m_{\v i}\})$:
\begin{eqnarray}
\Psi_T(\{ m_{\v i}\})=\sum_{ijkl\cdots}
T_{\v i,ejif}^{m_1}T_{\v j,jhkg}^{m_2}T_{\v k,lqrk}^{m_3}T_{\v
l,tlsi}^{m_4}\cdots
\label{TNS}
\end{eqnarray}
Here $T_{\v i,ejfi}^{m_{\v i}}$ is a complex tensor on
vertex $\v i$ with one physical index $m_{\v i}$ and four
inner indices $i,j,k,l,\cdots$. The physical index runs over
the number of physical states $D_p$ on each vertex, and the
internal indices runs over $D$ values.  Note that tensors on
different vertices can be different.  The TNS can be
represented graphically as in Fig. \ref{tps}.

We see that a TNS is characterized by tensors
$T_{\v i,ijkl}^{m}$ and by a network specifying the connectivity.  
Can we calculate the topological properties
of the many body state $\Psi_T(\{ m_{\v i}\})$ from its
defining tensors $T_{\v i,ijkl}^{m}$? To answer this question, we
need to define the problem more completely by introducing a Hamiltonian
$H_T$ such that $\Psi_T(\{ m_{\v i}\})$ is the exact ground
state of $H_T$.  We will call $H_T$ an ``ideal'' Hamiltonian
and $\Psi_T(\{ m_{\v i}\})$ an ``ideal'' wave function.

\begin{figure}
\begin{center}
\includegraphics[scale=0.5]{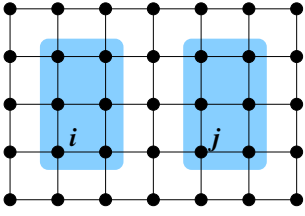}
\end{center}
%Fig 2
\caption{
\label{slatt}
(Color online)
The patch $P_{\v i}$ and the patch $P_{\v j}$.  $P_{\v i}$
and $P_{\v j}$  only differ by a displacement.
}
\end{figure}

The Hamiltonian $H_T$ has the following local form
\begin{align}
 H_T=\sum_{\v i} O_{P_{\v i}}
\end{align}
where $O_{P_{\v i}}$ is an operator that acts only on states in 
the patch $P_{\v i}$ (see Fig. \ref{slatt}).  How can we 
choose the operator $O_{P_{\v i}}$ so that $\Psi_T(\{
m_{\v i}\})$ is, hopefully, the only ground state of $H_T$?
In the following, we will describe a construction
proposed in \Ref{PVC0850}.

Let $\rho_{P_i}$ be the density matrix of the state
$\Psi_T(\{ m_{\v i}\})$ on patch $P_{\v i}$ obtained by tracing
out the physical degrees of freedom outside $P_{\v i}$:
\begin{align}
 \rho_{P_i}=\Tr_{\bar P_{\v i}} |\Psi_T\>\<\Psi_T|
\end{align}
where $\bar P_{\v i}$ represent the region outside of $P_{\v
i}$ and $ \Tr_A$ is the trace over all the states in region
$A$.  From the structure of the TNS, we can deduce that
$\rho_{P_i}$ must have the form
\begin{align}
 \rho_{P_i}=\sum_{a=1}^{D^{2L_{P_{\v i}}} } |\psi_a\>\<\phi_a|,
\end{align}
where $|\psi_a\>$ and $|\phi_a\>$ are states on the patch
$P_{\v i}$, and $L_{P_{\v i}}$ is the number of links on the
boundary of the patch $P_{\v i}$ (for example $L_{P_{\v
i}}=10$ for the patch in Fig.  \ref{slatt}).  Equivalently,
$\rho_{P_i}$ must have the form
\begin{align}
 \rho_{P_i}=\sum_{a=1}^{ N_{P_{\v i}}} \rho_a |\vphi_a\>\<\vphi_a|,
.\ \ \ \ N_{P_{\v i}} \leq D^{2L_{P_{\v i}}} ,
\end{align}
where $|\vphi_a\>$ are normalized states on the patch $P_{\v
i}$.  This means that $\rho_{P_i}$ can be constructed from only
$N_{P_{\v i}} \leq D^{2L_{P_{\v i}}}$ states on patch $P_{\v i}$.  Since the
number of states on $P_{\v i}$ grows like $D_p^{\al L_{P_{\v i}}^2},\
\al= O(1)$ for a large patch (here $d_\text{space}=2$), the following projection
operator
\begin{align}
P_{P_{\v i}}=\sum_{a=1}^{N_{P_{\v i}}}  |\vphi_a\>\<\vphi_a|
\end{align}
is non-trivial in the large patch limit (\ie $1-P_{P_{\v i}}
\neq 0$) because $N_{P_{\v i}} \leq D^{2L_{P_{\v i}}} < D_p^{\al
L_{P_{\v i}}^2}$.  If we choose
\begin{align}
\label{HTP}
 H_T=\sum_{\v i} (1-P_{P_{\v i}}) ,
\end{align}
the TNS $\Psi_T$ will be the exact ground state
of $H_T$. We hope that in the large patch limit,
the projection $P_{P_{\v i}}$ is highly restrictive
and $\Psi_T$ is more or less the only ground state.

Now the problem can be better defined. From the tensors
$T^m_{\v i,ijkl}$, we can determine an ``ideal'' Hamiltonian
$H_T$ \eq{HTP}.  Assuming $H_T$ is gapped, the topological
properties of the TNS are just those of the topological
phase of $H_T$.  However, in this paper we will not try to
calculate the topological properties by directly
diagonalizing $H_T$.  This task is generically difficult
because $ [ P_{P_{\v i}}, P_{P_{\v j}}]\neq 0$.  Instead, we
will find a way to calculate the topological properties from
the structure of the tensors $T^m_{\v i,ijkl}$ directly.

\section{Local Gauge Structure of a TNS}

\begin{figure}
\begin{center}
\includegraphics[scale=0.6]{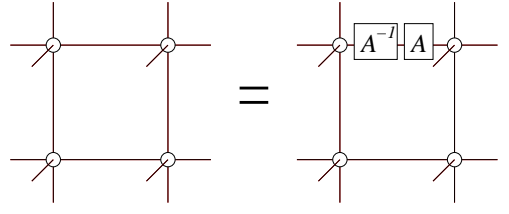}
\end{center}
%Fig 3
\caption{
\label{gauge}
A graphical representation of a gauge transformation.
}
\end{figure}

\begin{figure}
\begin{center}
\includegraphics[scale=0.5]{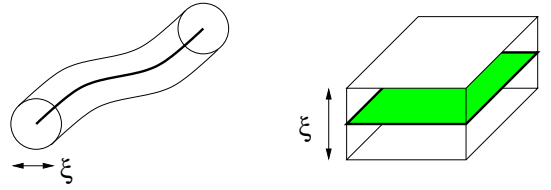}
\end{center}
%Fig 4
\caption{
\label{brane}
(Color online)
A 1-brane and a 2-brane.
}
\end{figure}

One of the important properties of a TNS is that different
tensors $T^m_{\v i,ijkl}$ can give rise to the same physical
wave function $\Psi_T(\{m_{\v i}\})$.  To be more precise,
two sets of tensors $T^m_{\v i,ijkl}$ and $\t T^m_{\v
i,ijkl}$ related by a ``gauge transformation'' (GT)
\begin{align}
\label{AAAAT}
&\ \ \  \La_{\v i}\t T^m_{\v i,l' r' d' u'}
\\
&= \sum_{lrud}
(A_{\v i-\v x,\v i})_{l'l}
(A_{\v i+\v x,\v i})_{r'r}
(A_{\v i+\v y,\v i})_{u'u}
(A_{\v i-\v y,\v i})_{d'd}
 T^m_{\v i,l r d u} ,
\nonumber \\
\label{AA}
& \sum_k (A_{\v j,\v i})_{kj}
(A_{\v i,\v j})_{ki}=\La_{\v i\v j} \del_{ij}
\end{align}
give rise to the same wave function $\Psi_T(\{m_{\v i}\})$
up to some scaling factors (see Fig. \ref{gauge}).
In the above $\La_{\v i}$ and
$\La_{\v i\v j}$ are scaling factors that do not depend on
the indices $i,j,k,...$.  Here $A_{\v i,\v j}$ and $A_{\v
j,\v i}$ are $D\times D$ invertible matrices defined on the
links $\<\v i,\v j\>$.  A GT is called a ``$d$-dimensional gauge
transformation'' ($d$-GT)  if $A_{\v j,\v i} \neq 1$ only on
links within with a ``$d$-brane''.  By ``$d$-brane'' we mean a
region near a $d$-dimensional subspace (\ie a
$d$-dimensional subspace with a finite ``thickness'' $\xi$
as illustrated in Fig. \ref{brane}).  Note that the
``thickness'' $\xi$ is fixed for a TNS, regardless the size
of the $d$-dimensional subspace and the value of $d$.
The size $L$ of the $d$-dimensional subspace is typically much bigger than
$\xi$: $\xi/L\to 0$.

Using $d$-GT we can introduce the concept of a
``$d$-dimensional full invariant gauge group'' ($d$-fIGG) as
in \Ref{Wqoslpub}.  $d$-fIGG is a property of the tensors
$T^m_{\v i,ijkl}$ and a fixed $d$-brane.  $d$-fIGG is a
group formed by all the $d$-GT in the $d$-brane that leave
$T^m_{\v i,ijkl}$ invariant up to a scaling factor.

We note that if $d$-fIGG is non trivial then we can use
$d$-GT in $d$-fIGG with different $d$-branes
to construct a non-trivial $d'$-fIGG for
all $d'>d$.  Thus, to reveal the new gauge structure that appears
at every dimension $d$, we will introduce the ``$d$-dimensional
invariant gauge group'' ($d$-IGG).  The $d$-fIGG contains a normal
subgroup whose elements are in $(d-1)$-fIGG. The $d$-IGG
is defined as
\begin{align}
\label{p-IGG}
 d\text{-IGG}\equiv d\text{-fIGG}/(d-1)\text{-fIGG}
\end{align}
The $d$-IGG for the tensors
$T^m_{\v i,ijkl}$ defined this way may depend on the
topology of the $d$-brane.

As an example of this notation, we note that for a
$d_\text{space}$-dimensional TNS, the $d_\text{space}$-IGG
is like the group of uniform global gauge transformations in
a gauge theory.

\section{Local Symmetry Structure of a TNS}

In the previous section we discussed gauge transformations
that leave the tensors of the TNS invariant (up to a scaling
factor).  These gauge transformations define the $d$-fIGG.
In this section, we will discuss gauge transformations
that do change the tensors $T^m_{\v i,ijkl}$.

Let us first introduce some concepts.  A ``local physical
transformation'' of a TNS is generated by unitary $D_p\times D_p$
matrices $S_{\v i,mm'}$ acting on the physical indices $m$
(here $D_p$ is the range of the
physical index $m$):
\begin{align}
 T^m_{\v i,ijkl} \to \sum_{m'} S_{\v i,mm'} T^{m'}_{\v i,ijkl}
\end{align}
A ``$d$-dimensional local physical transformation''
($d$-lPT) is a ``local physical transformation'' that is
non-trivial (\ie $ S_{\v i,mm'}\neq \del_{mm'}$) only on a
$d$-brane.  A ``$d$-dimensional symmetry transformation''
($d$-ST) of a TNS is a $d$-lPT that leave the TNS invariant.
All the $d$-ST on a fixed $d$-brane from a group which will
be called the ``$d$-dimensional full symmetry group''
$d$-fSG.

The $d$-fSG is very similar to the ``low-dimensional
gauge-like symmetries'' introduced in \Ref{NO0616}.  A
difference between the two concepts is that $d$-fSG  is a
symmetry of a ground state while the ``low-dimensional
gauge-like symmetries'' are symmetries of an exactly solvable
Hamiltonian.  As pointed in \Ref{NO0616}, the
low-dimensional gauge-like symmetries are useful tools to
calculate the topological properties of a model
Hamiltonian.  Similarly, $d$-fSG should be a useful tool to
calculate topological properties from a ground state.

We note that if $d$-fSG is non trivial, then we can use
elements in $d$-fSG on different $d$-branes to construct a
non-trivial $d'$-fSG for all $d'>d$.  Thus, to reveal the
new structure that appears in each dimension $d$, we
introduce the ``$d$-dimensional symmetry group'' $d$-SG.  As
in the previous section, the $d$-fSG contains a normal
subgroup that is formed by elements in $(d-1)$-fSG. We
define $d$-SG as
\begin{align}
\label{p-SG}
 d\text{-SG}\equiv d\text{-fSG}/(d-1)\text{-fSG}
\end{align}

We can make contact with the string-net framework by noticing that
 the elements of $1$-SG are actually the ideal string operators
\cite{K032,Wqoexct,LWstrnet} that satisfy a ``zero
law''.\cite{HW0541} Similarly, the elements of $d$-SG are the
ideal $d$-brane operators that satisfy a ``zero
law''.\cite{BM0636,HZW0507} So $d$-SG and
``low-dimensional gauge-like symmetries'' are closely
related to the string operators and the $d$-brane operators
that play a key role in topologically ordered
states.\cite{FNS0428,LWsta,LWstrnet}

The above discussion of $d$-SG (or ``low-dimensional
gauge-like symmetries'') is for a generic topological state.
Now the question is how to calculate $d$-SG for a TNS?
Naively, we would like to say that $d$-SG are formed by local
physical transformations that leave the tensors $T^m_{\v
i,ijkl}$ invariant.  But due to the gauge structure of the
TNS, the invariance of the TNS and the invariance of the
tensors $T^m_{\v i,ijkl}$ are not equivalent.  Instead, $d$-SG
are formed by local physical transformations that leave
the tensors $T^m_{\v i,ijkl}$ invariant up to gauge
transformations.  Thus the gauge transformations that change
the tensors play an important role in calculating $d$-SG for
a TNS.

In the above, we have defined a notion of $d$-IGG and $d$-SG for a TNS.
We will see that both $d$-IGG and $d$-SG play a very important
role in characterizing a TNS.  Many (if not all) topological
properties of a TNS can be calculated from its $d$-IGG and
$d$-SG.  In the following, we will carry out the this
program for the simplest TNS with a non-trivial topological
order.

\section{$Z_2$ Topologically Ordered State}

\subsection{$Z_2$ String Condensed State and its Ideal Hamiltonian}

\begin{figure}[tb]
\centerline{
\includegraphics[scale=0.6]{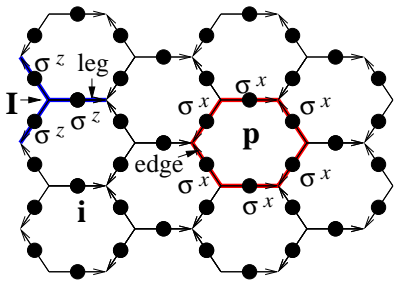}
}
%Fig 5
\caption{
(Color on line)
A Kagome lattice viewed as the links of a honeycomb lattice.
}
\label{KlattZ2}
\end{figure}

The simplest topological phase that we can consider is the
$Z_2$ topological phase.\cite{RS9173,Wsrvb}  Such a
phase can be realized through a spin-1/2 model on the Kagome
lattice:\cite{K032,LWstrnet}
\begin{align}
\label{Hz2}
H &= U \sum_{\v I} (1-Q_{\v I})  + g \sum_{\v p} (1-B_{\v p})
\nonumber\\
Q_{\v I} &=\prod_{\text{legs of }\v I} \si^z_{\v i},\ \ \ \
B_{\v p} =\prod_{\text{edges of }\v p} \si^x_{\v i}.
\end{align}
Here we have viewed the Kagome lattice as the links of the honeycomb
lattice (see Fig. \ref{KlattZ2}).  The vertices of the honeycomb
lattice are labeled by $\v I$, the links (which are the sites of the
Kagome lattice) by $\v i$, and the faces by $\v p$.  $\sum_{\v I}$
sums over all vertices and $\sum_{\v p}$ over all faces.
The Hamiltonian \eq{Hz2} indeed has the form of a summation of
projectors as in \eqn{HTP}.

\begin{figure}[tb]
\centerline{
\includegraphics[scale=0.55]{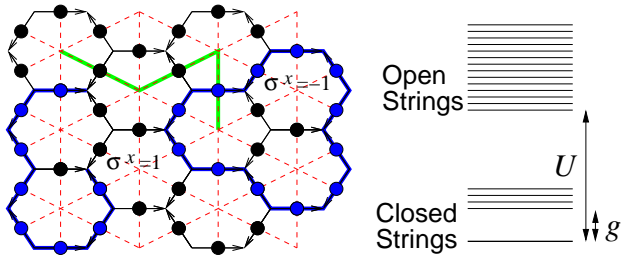}
}
%Fig 6
\caption{
(Color online)
A spin configuration of $\si^z=1$ (filled dot) and
$\si^z=-1$ (open dot) on the Kagome lattice can be viewed as a
string configuration on the honeycomb lattice.  The end of
an open string costs an energy $U$.  The closed-string
sector has an energy gap $g$.  The dashed lines describe the
dual lattice of the honeycomb lattice.  The thick blue lines
are strings in the honeycomb lattice.  The thick green
lines are strings in the dual lattice.
}
\label{KlattLp}
\end{figure}

We can interpret a given configuration of spins in terms of strings by
viewing the state $\sigma^z = 1$ as the absence of a string and the
state $\sigma^z = -1$ as the presence of a string (note the choice of
$\sigma^z$ here).  The above Hamiltonian has the property that, in the
large $U>0$ limit, the low energy Hilbert space is the space of all
spin configurations that contain only closed strings (see Figure
\ref{KlattLp}).
For $g>0$, the ground state of $H$ is the equal weighted superposition of all
closed string states $|\Omega \rangle = \sum_{X \text{closed}}
|X\rangle$.  It is useful to consider string operators that add a string to the system.  These operators can be taken to be
\begin{align}
\label{XC}
X(C) = \prod_{i \in C} \sigma^x_{\v i}
\end{align}
where $C$ is a curve running along the
edges of the lattice.  The operator $X(C)$ is a string creation operator
because $\sigma^x$ acts as a spin flip operator for eigenstates of
$\sigma^z=m$ so that we are simply flipping the spins (adding a string)
along the curve $C$.  Note that the operator $X(C)$ can also remove a
string because the string is its own anti-string meaning that two
strings can annihilate each other.

The ground state $|\Omega\rangle$ has the remarkable feature
that it is an eigenstate of $X(C)$ for all $C$ with
eigenvalue $1$.  We say that the strings created by $X(C)$
have condensed in the state $|\Omega\rangle$.  Another
useful point of view comes from thinking about the $Z_2$
condensed state in terms of $Z_2$ gauge theory where the
operator $X(C)$ is nothing but a Wilson-Wegner loop
operator.  In the string condensed or deconfined phase the
Wilson-Wegner loop $X(C)$ satisfies a ``zero law'' as
opposed to an area or perimeter law.\cite{HW0541} This
``zero law'' expresses the existence of strings at all
scales in the ground state.

For an open string operator that satisfies the ``zero law'',
the action of the string operator will create a pair of
excitations at its two ends.  Such excitations will in
general have fractional statistics and fractional quantum
numbers.  For the $X(C)$ string, the excitations at its ends
correspond to $Z_2$ charge. Thus we will call the $X(C)$ string
an electric string.

There is another string operator that we can consider, which we define
as
\begin{align}
\label{ZC}
Z(C^*) = \prod_{i \in C^*} \sigma^z_{\v i}
\end{align}
where $C^*$ is a curve
along the links of the dual lattice (see Figure \ref{KlattLp}).
It is characteristic of the $Z_2$ string condensed
phase that these strings are also condensed: $Z(C^*)|\Omega
\rangle = | \Omega \rangle$ for any loop $C^*$.

For the $Z(C^*)$ string, the excitations at its ends
correspond to $Z_2$ vortices. Thus we will call the $Z(C^*)$
string a magnetic string.

This was all for the case of a system on an infinite plane where there
is a unique ground state.  Things change qualitatively when the system
lives on a manifold with non-trivial topology.  When the manifold in
question has non-trivial $1$-cycles (non-contractible loops) then the
string operators corresponding to these non-contractible loops become
interesting observables.  We will return to the case of non-trivial
topology later.

\subsection{$Z_2$ Tensor Network Representation}

\begin{figure}[tb]
\begin{center} \includegraphics[width=2.9in] {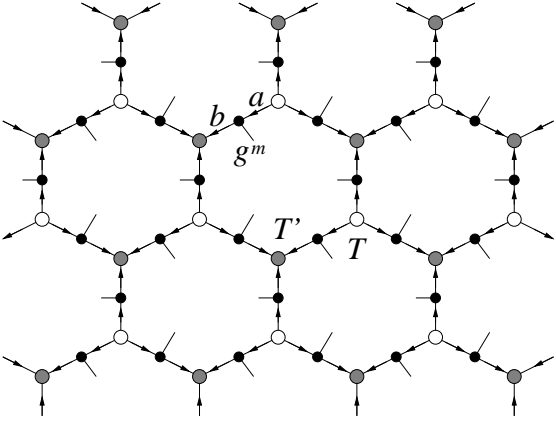}
%Fig 7
\end{center}
\caption{
A tensor network formed by vertices and links.  The links that
connect the dots carry indices $a,b,...$.  Each trivalent vertex
represents a $T$-tensor.  The vertices on the A-sublattice (open dots
labeled by $\v I_A$) represent $T_{abc}$. The vertices on the
B-sublattice (shaded dots $\v I_B$) represent $T'_{abc}$.
The dots on the links represent the $g^m$-tensor
$g^m_{ab}$.  In the tensor trace $\text{tTr}\otimes_{\v I_A} T
\otimes_{\v I_B} T' \otimes_{\v i} g^{m_{\v i}}$, the $a,b,c,...$
indices on the links that connect the dots are summed over.
The tensor trace produces a complex number
$\Phi(m_1,m_2,...)$ that depends on $m_{\v i}$ which can be viewed as
a wave function.
}
\label{hexgT}
\end{figure}

\begin{figure}[tb]
\begin{center}
\includegraphics[scale=0.6] {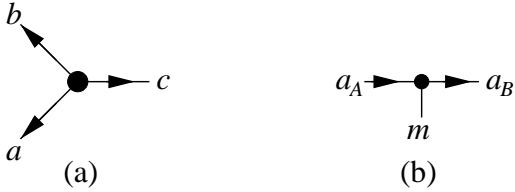}
%Fig 8
\end{center}
\caption{
 The graphic representation of (a) the $T$-tensor,
$T_{abc} $, and (b) the $g^m$-tensor $ g^m_{a_Ab_B}
$.
} \label{gT}
\end{figure}

The $Z_2$ condensed state, and many other topologically
ordered states, can be systematically represented in terms
of a TNS.\cite{BAV0919,GLS0918} Fig. \ref{hexgT}
and \ref{gT} represent the tensor network on the honeycomb
lattice graphically.  Note that the tensor network is formed
by three tensors $T_{abc}$, $T'_{abc}$, and $g^m_{ab}$.
To describe the $Z_2$ condensed state, we choose
$D=2$ and
\begin{align}
& g^{m=1}_{00}=1, \ \ \
 g^{m=-1}_{11}=1, \ \ \ \text{other } g^m_{ab}=0;
\nonumber\\
& T_{abc}=T'_{abc}=1 \text{ if } a+b+c=\text{even},
\nonumber\\
& T_{abc}=T'_{abc}=0 \text{ if } a+b+c=\text{odd}.
\end{align}
where $a,b,c=0,1$.

In the following, we will calculate the string operators
directly from the above tensors using the $d$-IGG and
$d$-SG of the tensor network.  This allows us to understand
the topological properties of the TNS directly from its defining
tensors.

\subsection{$d$-IGG of the $Z_2$ Tensor Network}

What is the $d$-IGG of the above $Z_2$ tensor network?
First, we find that the
$0$-IGG is formed by the following gauge transformations:
\begin{align}
 (A_{\v i\v I})_{ab}=\la_{\v i\v I}\del_{ab},\ \ \ \
 (A_{\v I\v i})_{ab}=\la_{\v I\v i}\del_{ab}.
\end{align}
These transformations only change the tensors
$g$, $T$ and $T'$ by a scaling factor.

Next, let us calculate $2$-IGG.
We consider the ``gauge transformations''
that leave $g^m_{ab}$ on link-$\v i$ invariant, up to
scaling factor:
\begin{align}
\La_{\v i}g^m_{a'b'}=
\sum_{ab}
(A_{\v I\v i})_{a'a}
(A_{\v J\v i})_{b'b}
g^m_{ab},\ \ \ \text{for } m=\pm 1.
\end{align}
Note that $\v I$ and $\v J$ label the vertices of the
honeycomb lattice and $\v i$ is the link that connect the
two vertices $\v I$ and $\v J$.  The above equation requires that
$A_{\v I\v i})$ and $A_{\v J\v i})$ be diagonal.  Using
the ``local gauge transformation'':
\begin{align}
g^m_{ab}\to
\sum_{ab}
\la_1\del_{a'a}\,
\la_2 \del_{b'b}
g^m_{ab},
\end{align}
we can fix $(A_{\v I\v i})_{00}= (A_{\v J\v i})_{00}=1$ and
\begin{align}
 (A_{\v I\v i})_{11} (A_{\v J\v i})_{11}=1 .
\end{align}
From \eqn{AA}, we find that
$ A_{\v i\v I}$ must be also diagonal.
Using the 0-GT in 0-IGG=0-fIGG
we can transform the diagonal
$ A_{\v i\v I}$ into the following form
\begin{align}
\label{AinvA}
 (A_{\v i\v I})_{00} =1, \ \ \ \
 (A_{\v i\v I})_{11} =1/(A_{\v I\v i})_{11} .
\end{align}
$ A_{\v i\v I}$ must leave $T$ invariant and thus
\begin{align}
 T_{a'b'c'}=\sum_{abc}
(A_{\v i\v I})_{a'a}
(A_{\v j\v I})_{b'b}
(A_{\v k\v I})_{c'c}
 T_{abc},
\end{align}
where $\v i$, $\v j$, and $\v k$ are the three links that
connect to the vertex $\v I$.
The above gives us
\begin{align}
(A_{\v i\v I})_{11} (A_{\v j\v I})_{11}=
(A_{\v j\v I})_{11} (A_{\v k\v I})_{11}=
(A_{\v k\v I})_{11} (A_{\v i\v I})_{11}=1.
\end{align}
Such a non-linear equation gives two sets of solutions
\begin{align}
 A_{\v i\v I}=A_{\v I\v i}=1,
\end{align}
and
\begin{align}
\label{Z2trans}
 A_{\v i\v I}=A_{\v I\v i}=\si^z.
\end{align}
where we have used \eqn{AinvA}.
So $2$-IGG of the $Z_2$ tensor network contains only two elements
given by the above two expressions.  Such a $2$-IGG is a $Z_2$
group.

Based on this discussion, we see that we can obtain the 2-IGG
from the 2-fIGG by removing the 0-GT, \ie
$2\text{-IGG}=2\text{-fIGG}/0\text{-fIGG}$.  This implies that the
1-IGG is the trivial group containing only the identity.

\subsection{Ground State Degeneracy from $d$-IGG}

Using the $d$-IGG of the TNS, we can calculate some topological
properties of the TNS.  In the $Z_2$ condensed state we found that
2-IGG=$Z_2$. Now we will show that the ground state
degeneracy of the $Z_2$ condensed state on a torus is the same
as the ground state degeneracy of $Z_2$ gauge theory on a
torus.

\begin{figure}[tb]
\begin{center}
\includegraphics[scale=0.5] {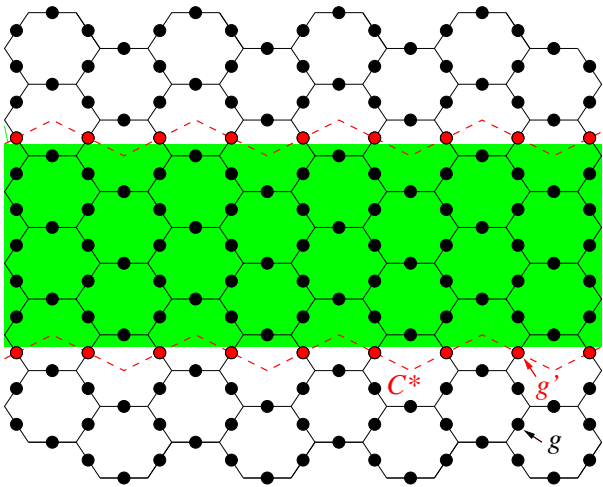}
%Fig 9
\end{center}
\caption{
(Color on line)
The $Z-2$ gauge transformation
 $A_{\v i\v I}=
A_{\v I\v i}=\si^z$ acts only on the shaded region.
The tensor $g$ is changed to $\t g$ on the boundary of the
stripe.
} \label{z2trns}
\end{figure}

If we apply the $Z_2$ gauge transformation $A_{\v i\v I}=
A_{\v I\v i}=\si^z$ every where, the tensors $T$, $T'$ and
$g$ are not changed.  If we apply the $Z_2$ gauge
transformation only on a stripe (see Fig. \ref{z2trns}),
then the tensors on the boundary of the stripe will be
changed.  For the particular choice of the stripe in Fig.
\ref{z2trns} the tensor $g$ is changed to $\t g$ on the
boundary of the stripe, where
\begin{align}
\t g^{m=1}_{00}=1, \ \ \
 \t g^{m=-1}_{11}=-1, \ \ \ \text{other } \t g^m_{ab}=0 .
\end{align}
We note that the change $g \to \t g$ along the boundary of
the stripe is generated by the $Z(C^*)$ string operator (see
\eqn{ZC}).  Therefore the $Z_2$ gauge transformation on a finite
region defines a type of string operator along the
boundary.

Since the application of the $Z_2$ gauge transformation
does not change the energy of the state, we find that the
application of the $Z(C^*)$ string operator also does not
change the energy of the state.  We note that the action of
two $Z(C^*)$ string operators along the top and the bottom
boundary of the stripe in Fig. \ref{z2trns} is equivalent to
a $Z_2$ gauge transformation on the stripe.  In other words,
the tensors are invariant under the action of the two
$Z(C^*)$ string operators, up to a $Z_2$ gauge
transformation on the stripe.  Thus the product of the two
$Z(C^*)$ string operators is an element in 1-SG.  The action
of the two $Z(C^*)$ string operators does not change the
state.

On the other hand the action of one $Z(C^*)$ string operator
is not equivalent to a $Z_2$ gauge transformation and is not
an elements of 1-SG.  Such an action will change the state,
but not the energy.  Applying such a $Z(C^*)$ string
operator to the $Z_2$ condensed state on torus along the
$x$- and $y$-directions will give us the four degenerate
ground state of the $Z_2$ condensed state on a torus.  This is
how do we calculate the ground state degeneracy from the
tensors in the TNS.

We would like to stress that the $Z(C^*)$ string
operator is directly obtained from the $Z_2$ gauge
transformation. The $Z_2$ group structure of the gauge group also
determines the $Z_2$ structure of the $Z(C^*)$ string
operator: $[Z(C^*)]^2=1$.

\subsection{$Z_2$ $d$-SG and the
Electric String Operator}

We have seen that the $d$-IGG of a TNS can be calculated by
finding all the gauge transformations that leave the tensors
invariant (up to a scaling factor).  Similarly, the $d$-SG
of a TNS can be calculated by finding all the combined local
physical transformations and the gauge transformations that
leave the tensors invariant.

For the case of the $Z_2$ condensed state, we have seen
that the magnetic string operator $Z(C^*)$ on a contractible
loop $C^*$ is a 1-ST (an element of 1-SG).  The tensors in
the tensor network are invariant under $Z(C^*)$ followed by
a 2-GT on the disk $D$ bounded by $C^*$.  The 1-SG generated
by such a 1-ST, $Z(C^*)$, will be called a non-local 1-SG.

There is another type of 1-ST, such that the tensors
in the tensor network are invariant under the 1-ST followed
by a 1-GT on the same curve on which the 1-ST is defined.
The 1-SG generated by this type of 1-ST will be called a local
1-SG.

The $Z_2$ condensed state also has a local 1-SG, which is generated
by the electric string operator $X(C)$ (see \eqn{XC}).  One
can directly check that the tensors $g$, $T$, and $T'$ on
the tensor network are invariant under the action of $X(C)$ followed
by a gauge transforation on the loop $C$ with
\begin{align}
 A_{\v i\v I}= A_{\v I\v i}= \si^x
\end{align}
on the loop.

We have seen that the ground state degeneracy on a torus can be
calculated from the string operators.  We can also calculate
the quasiparticle statistics from the string operators.
This is because an open string operator creates a pair of
quasiparticles at its ends.  The electric string operator
$X(C)$ creates a pair of $Z_2$ charges, and the magnetic
string operator $Z(C^*)$ creates a pair of $Z_2$ vortices.
The open string operators can also be viewed as hopping
operators for the quasiparticles.  The statistics of the
quasiparticles can then be calculated from the algebra of
these hopping operators (\ie the open string operators).\cite{LWsta}

\section{Discussion and Conclusions}

In this paper, we viewed a TNS as an ideal wave function that
is the exact ground state of a ideal Hamiltonian.  When the
ideal Hamiltonian has a finite energy gap, the TNS can represent a
quantum phase with non-trivial topological order.  We argued that the
topological properties of such a phase can be calculated just from
the tensors that form the TNS.

Motivated by the gauge group and projective symmetry group
in the slave-particle/projective construction
\cite{Wqoslpub} and by low-dimensional gauge-like symmetries
of model Hamiltonians\cite{NO0616}, we introduced the
$d$-IGG and the $d$-SG for a TNS.  Using the $d$-IGG and the
$d$-SG of a TNS, we can calculate the string operators (or
$d$-brane operators) of the TNS, as we demonstrated in the
simple $Z_2$ condensed state.  Many topological properties
of the TNS, such as ground state degeneracy and
quasiparticle statistics can be calculated from the
resulting string operators.  This allows us to identify
the topological order of the TNS just from the tensors that
form the TNS.

While completing this paper, we became aware a preprint by
Schuch, Cirac, and Perez-Garcia (arXiv:1001.3807) where a
class of TNS satisfying the ``G-injective" condition are studied
(injectivity means that one can achieve any action on the
inner indices $a,b,c,...$ by acting on the physical indices
$m$). From the symmetry properties of the tensors
(similar to our $d_\text{space}$-IGG), they can also calculate
ground state degeneracy and other physical properties from
the tensors of the tensor network.

This research is supported by  NSF Grant No. DMR-0706078

%\bibliography{../../bib/wencross,../../bib/all,../../bib/publst}

\begin{thebibliography}{42}
\expandafter\ifx\csname natexlab\endcsname\relax\def\natexlab#1{#1}\fi
\expandafter\ifx\csname bibnamefont\endcsname\relax
  \def\bibnamefont#1{#1}\fi
\expandafter\ifx\csname bibfnamefont\endcsname\relax
  \def\bibfnamefont#1{#1}\fi
\expandafter\ifx\csname citenamefont\endcsname\relax
  \def\citenamefont#1{#1}\fi
\expandafter\ifx\csname url\endcsname\relax
  \def\url#1{\texttt{#1}}\fi
\expandafter\ifx\csname urlprefix\endcsname\relax\def\urlprefix{URL }\fi
\providecommand{\bibinfo}[2]{#2}
\providecommand{\eprint}[2][]{\url{#2}}

\bibitem[{\citenamefont{Wen}(1989)}]{Wtop}
\bibinfo{author}{\bibfnamefont{X.-G.} \bibnamefont{Wen}},
  \bibinfo{journal}{Phys. Rev. B} \textbf{\bibinfo{volume}{40}},
  \bibinfo{pages}{7387} (\bibinfo{year}{1989}).

\bibitem[{\citenamefont{Wen}(1990)}]{Wrig}
\bibinfo{author}{\bibfnamefont{X.-G.} \bibnamefont{Wen}},
  \bibinfo{journal}{Int. J. Mod. Phys. B} \textbf{\bibinfo{volume}{4}},
  \bibinfo{pages}{239} (\bibinfo{year}{1990}).
  
\bibitem[{\citenamefont{Wen}(1995)}]{Wtoprev}
\bibinfo{author}{\bibfnamefont{X.-G.} \bibnamefont{Wen}},
  \bibinfo{journal}{Advances in Physics} \textbf{\bibinfo{volume}{44}},
  \bibinfo{pages}{405} (\bibinfo{year}{1995}).

\bibitem[{\citenamefont{Kalmeyer and Laughlin}(1987)}]{KL8795}
\bibinfo{author}{\bibfnamefont{V.}~\bibnamefont{Kalmeyer}} \bibnamefont{and}
  \bibinfo{author}{\bibfnamefont{R.~B.} \bibnamefont{Laughlin}},
  \bibinfo{journal}{Phys. Rev. Lett.} \textbf{\bibinfo{volume}{59}},
  \bibinfo{pages}{2095} (\bibinfo{year}{1987}).

\bibitem[{\citenamefont{Wen et~al.}(1989)\citenamefont{Wen, Wilczek, and
  Zee}}]{WWZcsp}
\bibinfo{author}{\bibfnamefont{X.-G.} \bibnamefont{Wen}},
  \bibinfo{author}{\bibfnamefont{F.}~\bibnamefont{Wilczek}}, \bibnamefont{and}
  \bibinfo{author}{\bibfnamefont{A.}~\bibnamefont{Zee}},
  \bibinfo{journal}{Phys. Rev. B} \textbf{\bibinfo{volume}{39}},
  \bibinfo{pages}{11413} (\bibinfo{year}{1989}).

\bibitem[{\citenamefont{Moore and Read}(1991)}]{MR9162}
\bibinfo{author}{\bibfnamefont{G.}~\bibnamefont{Moore}} \bibnamefont{and}
  \bibinfo{author}{\bibfnamefont{N.}~\bibnamefont{Read}},
  \bibinfo{journal}{Nucl. Phys. B} \textbf{\bibinfo{volume}{360}},
  \bibinfo{pages}{362} (\bibinfo{year}{1991}).
  
\bibitem[{\citenamefont{Wen}(1991{\natexlab{a}})}]{Wnab}
\bibinfo{author}{\bibfnamefont{X.-G.} \bibnamefont{Wen}},
  \bibinfo{journal}{Phys. Rev. Lett.} \textbf{\bibinfo{volume}{66}},
  \bibinfo{pages}{802} (\bibinfo{year}{1991}{\natexlab{a}}).

\bibitem[{\citenamefont{Read and Sachdev}(1991)}]{RS9173}
\bibinfo{author}{\bibfnamefont{N.}~\bibnamefont{Read}} \bibnamefont{and}
  \bibinfo{author}{\bibfnamefont{S.}~\bibnamefont{Sachdev}},
  \bibinfo{journal}{Phys. Rev. Lett.} \textbf{\bibinfo{volume}{66}},
  \bibinfo{pages}{1773} (\bibinfo{year}{1991}).

\bibitem[{\citenamefont{Wen}(1991{\natexlab{b}})}]{Wsrvb}
\bibinfo{author}{\bibfnamefont{X.-G.} \bibnamefont{Wen}},
  \bibinfo{journal}{Phys. Rev. B} \textbf{\bibinfo{volume}{44}},
  \bibinfo{pages}{2664} (\bibinfo{year}{1991}{\natexlab{b}}).  

\bibitem[{\citenamefont{Misguich et~al.}(1999)\citenamefont{Misguich,
  Lhuillier, Bernu, and Waldtmann}}]{MLB9964}
\bibinfo{author}{\bibfnamefont{G.}~\bibnamefont{Misguich}},
  \bibinfo{author}{\bibfnamefont{C.}~\bibnamefont{Lhuillier}},
  \bibinfo{author}{\bibfnamefont{B.}~\bibnamefont{Bernu}}, \bibnamefont{and}
  \bibinfo{author}{\bibfnamefont{C.}~\bibnamefont{Waldtmann}},
  \bibinfo{journal}{Phys. Rev. B} \textbf{\bibinfo{volume}{60}},
  \bibinfo{pages}{1064} (\bibinfo{year}{1999}).
  
\bibitem[{\citenamefont{Misguich et~al.}(2002)\citenamefont{Misguich, Serban,
  and Pasquier}}]{MSP0202}
\bibinfo{author}{\bibfnamefont{G.}~\bibnamefont{Misguich}},
  \bibinfo{author}{\bibfnamefont{D.}~\bibnamefont{Serban}}, \bibnamefont{and}
  \bibinfo{author}{\bibfnamefont{V.}~\bibnamefont{Pasquier}},
  \bibinfo{journal}{Phys. Rev. Lett.} \textbf{\bibinfo{volume}{89}},
  \bibinfo{pages}{137202} (\bibinfo{year}{2002}).  

\bibitem[{\citenamefont{Kitaev}(2003)}]{K032}
\bibinfo{author}{\bibfnamefont{A.~Y.} \bibnamefont{Kitaev}},
  \bibinfo{journal}{Ann. Phys. (N.Y.)} \textbf{\bibinfo{volume}{303}},
  \bibinfo{pages}{2} (\bibinfo{year}{2003}).

\bibitem[{\citenamefont{Moessner and Sondhi}(2003)}]{MS0312}
\bibinfo{author}{\bibfnamefont{R.}~\bibnamefont{Moessner}} \bibnamefont{and}
  \bibinfo{author}{\bibfnamefont{S.~L.} \bibnamefont{Sondhi}},
  \bibinfo{journal}{Phys. Rev. B} \textbf{\bibinfo{volume}{68}},
  \bibinfo{pages}{184512} (\bibinfo{year}{2003}).
  
\bibitem[{\citenamefont{Levin and Wen}(2003)}]{LWsta}
\bibinfo{author}{\bibfnamefont{M.}~\bibnamefont{Levin}} \bibnamefont{and}
  \bibinfo{author}{\bibfnamefont{X.-G.} \bibnamefont{Wen}},
  \bibinfo{journal}{Phys. Rev. B} \textbf{\bibinfo{volume}{67}},
  \bibinfo{pages}{245316} (\bibinfo{year}{2003}).  

\bibitem[{\citenamefont{Freedman et~al.}(2004)\citenamefont{Freedman, Nayak,
  Shtengel, Walker, and Wang}}]{FNS0428}
\bibinfo{author}{\bibfnamefont{M.}~\bibnamefont{Freedman}},
  \bibinfo{author}{\bibfnamefont{C.}~\bibnamefont{Nayak}},
  \bibinfo{author}{\bibfnamefont{K.}~\bibnamefont{Shtengel}},
  \bibinfo{author}{\bibfnamefont{K.}~\bibnamefont{Walker}}, \bibnamefont{and}
  \bibinfo{author}{\bibfnamefont{Z.}~\bibnamefont{Wang}},
  \bibinfo{journal}{Ann. Phys. (NY)} \textbf{\bibinfo{volume}{310}},
  \bibinfo{pages}{428} (\bibinfo{year}{2004}).

\bibitem[{\citenamefont{Levin and Wen}(2005)}]{LWstrnet}
\bibinfo{author}{\bibfnamefont{M.}~\bibnamefont{Levin}} \bibnamefont{and}
  \bibinfo{author}{\bibfnamefont{X.-G.} \bibnamefont{Wen}},
  \bibinfo{journal}{Phys. Rev. B} \textbf{\bibinfo{volume}{71}},
  \bibinfo{pages}{045110} (\bibinfo{year}{2005}).

\bibitem[{\citenamefont{Nayak et~al.}(2008)\citenamefont{Nayak, Simon, Stern,
  Freedman, and Sarma}}]{NSS0883}
\bibinfo{author}{\bibfnamefont{C.}~\bibnamefont{Nayak}},
  \bibinfo{author}{\bibfnamefont{S.~H.} \bibnamefont{Simon}},
  \bibinfo{author}{\bibfnamefont{A.}~\bibnamefont{Stern}},
  \bibinfo{author}{\bibfnamefont{M.}~\bibnamefont{Freedman}}, \bibnamefont{and}
  \bibinfo{author}{\bibfnamefont{S.~D.} \bibnamefont{Sarma}},
  \bibinfo{journal}{Rev. Mod. Phys.} \textbf{\bibinfo{volume}{80}},
  \bibinfo{pages}{1083} (\bibinfo{year}{2008}), \eprint{arXiv:0707.1889}.

\bibitem[{\citenamefont{Chamon}(2005)}]{C0502}
\bibinfo{author}{\bibfnamefont{C.}~\bibnamefont{Chamon}},
  \bibinfo{journal}{Phys. Rev. Lett.} \textbf{\bibinfo{volume}{94}},
  \bibinfo{pages}{040402} (\bibinfo{year}{2005}).

\bibitem[{\citenamefont{Verstraete and Cirac}(2004)}]{VC0466}
\bibinfo{author}{\bibfnamefont{F.}~\bibnamefont{Verstraete}} \bibnamefont{and}
  \bibinfo{author}{\bibfnamefont{J.~I.} \bibnamefont{Cirac}}
  (\bibinfo{year}{2004}), \eprint{arXiv:cond-mat/0407066}.

\bibitem[{\citenamefont{Gendiar et~al.}(2003)\citenamefont{Gendiar, Maeshima,
  and Nishino}}]{GMN0391}
\bibinfo{author}{\bibfnamefont{A.}~\bibnamefont{Gendiar}},
  \bibinfo{author}{\bibfnamefont{N.}~\bibnamefont{Maeshima}}, \bibnamefont{and}
  \bibinfo{author}{\bibfnamefont{T.}~\bibnamefont{Nishino}},
  \bibinfo{journal}{Prog. Theor. Phys.} \textbf{\bibinfo{volume}{110}},
  \bibinfo{pages}{691} (\bibinfo{year}{2003}).

\bibitem[{\citenamefont{Nishio et~al.}(2004)\citenamefont{Nishio, Maeshima,
  Gendiar, and Nishino}}]{NMG0415}
\bibinfo{author}{\bibfnamefont{Y.}~\bibnamefont{Nishio}},
  \bibinfo{author}{\bibfnamefont{N.}~\bibnamefont{Maeshima}},
  \bibinfo{author}{\bibfnamefont{A.}~\bibnamefont{Gendiar}}, \bibnamefont{and}
  \bibinfo{author}{\bibfnamefont{T.}~\bibnamefont{Nishino}}
  (\bibinfo{year}{2004}), \eprint{arXiv:cond-mat/0401115}.

\bibitem[{\citenamefont{Maeshima}(2004)}]{M0460}
\bibinfo{author}{\bibfnamefont{N.}~\bibnamefont{Maeshima}},
  \bibinfo{journal}{J. Phys. Soc. Jpn.} \textbf{\bibinfo{volume}{73}},
  \bibinfo{pages}{60} (\bibinfo{year}{2004}).

\bibitem[{\citenamefont{Aguado and Vidal}(2008)}]{AV0804}
\bibinfo{author}{\bibfnamefont{M.}~\bibnamefont{Aguado}} \bibnamefont{and}
  \bibinfo{author}{\bibfnamefont{G.}~\bibnamefont{Vidal}},
  \bibinfo{journal}{Phys. Rev. Lett.} \textbf{\bibinfo{volume}{100}},
  \bibinfo{pages}{070404} (\bibinfo{year}{2008}).

\bibitem[{\citenamefont{Gu et~al.}(2009)\citenamefont{Gu, Levin, Swingle, and
  Wen}}]{GLS0918}
\bibinfo{author}{\bibfnamefont{Z.-C.} \bibnamefont{Gu}},
  \bibinfo{author}{\bibfnamefont{M.}~\bibnamefont{Levin}},
  \bibinfo{author}{\bibfnamefont{B.}~\bibnamefont{Swingle}}, \bibnamefont{and}
  \bibinfo{author}{\bibfnamefont{X.-G.} \bibnamefont{Wen}},
  \bibinfo{journal}{Phys. Rev. B} \textbf{\bibinfo{volume}{79}},
  \bibinfo{pages}{085118} (\bibinfo{year}{2009}), \eprint{arXiv:0809.2821}.

\bibitem[{\citenamefont{Buerschaper et~al.}(2009)\citenamefont{Buerschaper,
  Aguado, and Vidal}}]{BAV0919}
\bibinfo{author}{\bibfnamefont{O.}~\bibnamefont{Buerschaper}},
  \bibinfo{author}{\bibfnamefont{M.}~\bibnamefont{Aguado}}, \bibnamefont{and}
  \bibinfo{author}{\bibfnamefont{G.}~\bibnamefont{Vidal}},
  \bibinfo{journal}{Phys. Rev. B} \textbf{\bibinfo{volume}{79}},
  \bibinfo{pages}{085119} (\bibinfo{year}{2009}), \eprint{arXiv:0809.2393}.

\bibitem[{\citenamefont{Gu et~al.}(2008)\citenamefont{Gu, Levin, and
  Wen}}]{GLWtergV}
\bibinfo{author}{\bibfnamefont{Z.-C.} \bibnamefont{Gu}},
  \bibinfo{author}{\bibfnamefont{M.}~\bibnamefont{Levin}}, \bibnamefont{and}
  \bibinfo{author}{\bibfnamefont{X.-G.} \bibnamefont{Wen}},
  \bibinfo{journal}{Phys. Rev. B} \textbf{\bibinfo{volume}{78}},
  \bibinfo{pages}{205116} (\bibinfo{year}{2008}).

\bibitem[{\citenamefont{Jiang et~al.}(2008)\citenamefont{Jiang, Weng, and
  Xiang}}]{JWX0803}
\bibinfo{author}{\bibfnamefont{H.~C.} \bibnamefont{Jiang}},
  \bibinfo{author}{\bibfnamefont{Z.~Y.} \bibnamefont{Weng}}, \bibnamefont{and}
  \bibinfo{author}{\bibfnamefont{T.}~\bibnamefont{Xiang}},
  \bibinfo{journal}{Phys. Rev. Lett.} \textbf{\bibinfo{volume}{101}},
  \bibinfo{pages}{090603} (\bibinfo{year}{2008}).

\bibitem[{\citenamefont{Gu and Wen}(2009)}]{GWtefr}
\bibinfo{author}{\bibfnamefont{Z.-C.} \bibnamefont{Gu}} \bibnamefont{and}
  \bibinfo{author}{\bibfnamefont{X.-G.} \bibnamefont{Wen}},
  \bibinfo{journal}{Phys. Rev. B} \textbf{\bibinfo{volume}{80}},
  \bibinfo{pages}{155131} (\bibinfo{year}{2009}), \eprint{arXiv:0903.1069}.

\bibitem[{\citenamefont{Levin and Nave}(2007)}]{LN0701}
\bibinfo{author}{\bibfnamefont{M.}~\bibnamefont{Levin}} \bibnamefont{and}
  \bibinfo{author}{\bibfnamefont{C.~P.} \bibnamefont{Nave}},
  \bibinfo{journal}{Phys. Rev. Lett.} \textbf{\bibinfo{volume}{99}},
  \bibinfo{pages}{120601} (\bibinfo{year}{2007}).

\bibitem[{\citenamefont{Jordan et~al.}(2008)\citenamefont{Jordan, Orus, Vidal,
  Verstraete, and Cirac}}]{JOV0802}
\bibinfo{author}{\bibfnamefont{J.}~\bibnamefont{Jordan}},
  \bibinfo{author}{\bibfnamefont{R.}~\bibnamefont{Orus}},
  \bibinfo{author}{\bibfnamefont{G.}~\bibnamefont{Vidal}},
  \bibinfo{author}{\bibfnamefont{F.}~\bibnamefont{Verstraete}},
  \bibnamefont{and} \bibinfo{author}{\bibfnamefont{J.~I.} \bibnamefont{Cirac}},
  \bibinfo{journal}{Physical Review Letters} \textbf{\bibinfo{volume}{101}},
  \bibinfo{pages}{250602} (\bibinfo{year}{2008}).

\bibitem[{\citenamefont{Orus and Vidal}(2009)}]{OV0903}
\bibinfo{author}{\bibfnamefont{R.}~\bibnamefont{Orus}} \bibnamefont{and}
  \bibinfo{author}{\bibfnamefont{G.}~\bibnamefont{Vidal}},
  \bibinfo{journal}{Phys. Rev. B} \textbf{\bibinfo{volume}{80}},
  \bibinfo{pages}{094403} (\bibinfo{year}{2009}).

\bibitem[{\citenamefont{Wen and Niu}(1990)}]{WNtop}
\bibinfo{author}{\bibfnamefont{X.-G.} \bibnamefont{Wen}} \bibnamefont{and}
  \bibinfo{author}{\bibfnamefont{Q.}~\bibnamefont{Niu}},
  \bibinfo{journal}{Phys. Rev. B} \textbf{\bibinfo{volume}{41}},
  \bibinfo{pages}{9377} (\bibinfo{year}{1990}).

\bibitem[{\citenamefont{Laughlin}(1983)}]{L8395}
\bibinfo{author}{\bibfnamefont{R.~B.} \bibnamefont{Laughlin}},
  \bibinfo{journal}{Phys. Rev. Lett.} \textbf{\bibinfo{volume}{50}},
  \bibinfo{pages}{1395} (\bibinfo{year}{1983}).

\bibitem[{\citenamefont{Arovas et~al.}(1984)\citenamefont{Arovas, Schrieffer,
  and Wilczek}}]{ASW8422}
\bibinfo{author}{\bibfnamefont{D.}~\bibnamefont{Arovas}},
  \bibinfo{author}{\bibfnamefont{J.~R.} \bibnamefont{Schrieffer}},
  \bibnamefont{and} \bibinfo{author}{\bibfnamefont{F.}~\bibnamefont{Wilczek}},
  \bibinfo{journal}{Phys. Rev. Lett.} \textbf{\bibinfo{volume}{53}},
  \bibinfo{pages}{722} (\bibinfo{year}{1984}).

\bibitem[{\citenamefont{Wen}(2002)}]{Wqoslpub}
\bibinfo{author}{\bibfnamefont{X.-G.} \bibnamefont{Wen}},
  \bibinfo{journal}{Phys. Rev. B} \textbf{\bibinfo{volume}{65}},
  \bibinfo{pages}{165113} (\bibinfo{year}{2002}).

\bibitem[{\citenamefont{Baskaran and Anderson}(1988)}]{BA8880}
\bibinfo{author}{\bibfnamefont{G.}~\bibnamefont{Baskaran}} \bibnamefont{and}
  \bibinfo{author}{\bibfnamefont{P.~W.} \bibnamefont{Anderson}},
  \bibinfo{journal}{Phys. Rev. B} \textbf{\bibinfo{volume}{37}},
  \bibinfo{pages}{580} (\bibinfo{year}{1988}).

\bibitem[{\citenamefont{Nussinov and Ortiz}(2006)}]{NO0616}
\bibinfo{author}{\bibfnamefont{Z.}~\bibnamefont{Nussinov}} \bibnamefont{and}
  \bibinfo{author}{\bibfnamefont{G.}~\bibnamefont{Ortiz}}
  (\bibinfo{year}{2006}), \eprint{cond-mat/0605316}.

\bibitem[{\citenamefont{Perez-Garcia et~al.}(2008)\citenamefont{Perez-Garcia,
  Verstraete, Cirac, and Wolf}}]{PVC0850}
\bibinfo{author}{\bibfnamefont{D.}~\bibnamefont{Perez-Garcia}},
  \bibinfo{author}{\bibfnamefont{F.}~\bibnamefont{Verstraete}},
  \bibinfo{author}{\bibfnamefont{J.~I.} \bibnamefont{Cirac}}, \bibnamefont{and}
  \bibinfo{author}{\bibfnamefont{M.~M.} \bibnamefont{Wolf}},
  \bibinfo{journal}{Quant. Inf. Comp.} \textbf{\bibinfo{volume}{8}},
  \bibinfo{pages}{650} (\bibinfo{year}{2008}).

\bibitem[{\citenamefont{Wen}(2003)}]{Wqoexct}
\bibinfo{author}{\bibfnamefont{X.-G.} \bibnamefont{Wen}},
  \bibinfo{journal}{Phys. Rev. Lett.} \textbf{\bibinfo{volume}{90}},
  \bibinfo{pages}{016803} (\bibinfo{year}{2003}).

\bibitem[{\citenamefont{Hastings and Wen}(2005)}]{HW0541}
\bibinfo{author}{\bibfnamefont{M.~B.} \bibnamefont{Hastings}} \bibnamefont{and}
  \bibinfo{author}{\bibfnamefont{X.-G.} \bibnamefont{Wen}},
  \bibinfo{journal}{Phys. Rev. B} \textbf{\bibinfo{volume}{72}},
  \bibinfo{pages}{045141} (\bibinfo{year}{2005}).

\bibitem[{\citenamefont{Bombin and Martin-Delgado}(2006)}]{BM0636}
\bibinfo{author}{\bibfnamefont{H.}~\bibnamefont{Bombin}} \bibnamefont{and}
  \bibinfo{author}{\bibfnamefont{M.}~\bibnamefont{Martin-Delgado}}
  (\bibinfo{year}{2006}), \eprint{cond-mat/0607736}.

\bibitem[{\citenamefont{Hamma et~al.}(2005)\citenamefont{Hamma, Zanardi, and
  Wen}}]{HZW0507}
\bibinfo{author}{\bibfnamefont{A.}~\bibnamefont{Hamma}},
  \bibinfo{author}{\bibfnamefont{P.}~\bibnamefont{Zanardi}}, \bibnamefont{and}
  \bibinfo{author}{\bibfnamefont{X.~G.} \bibnamefont{Wen}},
  \bibinfo{journal}{Phys. Rev. B} \textbf{\bibinfo{volume}{72}},
  \bibinfo{pages}{035307} (\bibinfo{year}{2005}).

\end{thebibliography}

\end{document}